\documentclass[useAMS,usenatbib,letterpaper]{mn2e}

\usepackage{graphicx}
\usepackage[english]{babel}
\usepackage{bm}
\usepackage{amssymb}
\newcommand{\pF}{p_\mathrm{F}}
\newcommand{\vF}{v_\mathrm{F}}
\newcommand{\kF}{k_\mathrm{F}}
\newcommand{\ai}{a_\mathrm{i}}
\newcommand{\cs}{{c_\mathrm{s}}}
\newcommand{\ktf}{k_\mathrm{TF}}
\newcommand{\dm}{\Delta\mu}

\newcommand{\dvp}{\Delta\varphi^\prime}
\newcommand{\dmp}{\Delta\mu^\prime}
\newcommand{\Tp}{T_\mathrm{p}}
\newcommand{\op}{\omega_\mathrm{p}}
\newcommand{\kB}{k_\mathrm{B}}
\newcommand{\qBZ}{q_\mathrm{BZ}}
\newcommand{\xr}{x_\mathrm{r}}

\title[Ion thermal conductivity]{Thermal conductivity of ions in a neutron star envelope}
\author[A. I.  Chugunov, P. Haensel]
{A. I.  Chugunov$^1$\thanks{E-mail: andr.astro@mail.ioffe.ru (AIC);
haensel@camk.edu.pl (PH)} and
P. Haensel$^2$\footnotemark[1]\\
$^1$ Ioffe Physico-Technical Institute, Politekhnicheskaya 26,
194021 Saint-Petersburg, Russia \\
$^2$ N. Copernicus Astronomical Center, Polish
           Academy of Sciences, Bartycka 18, PL-00-716 Warszawa,
           Poland
           }

\begin{document}

\date{Accepted 2007 July 30. Received 2007 July 30;
in original form 2007 July 09}

\pagerange{\pageref{firstpage}--\pageref{lastpage}}
\pubyear{2007}

\maketitle

\label{firstpage}

\begin{abstract}
We analyze the thermal conductivity of ions (equivalent to the
conductivity of phonons in crystalline matter) in a neutron star
envelope.

We calculate the ion/phonon thermal conductivity in a crystal of
atomic nuclei using variational formalism and performing
momentum-space integration by Monte Carlo method. We take into
account phonon-phonon and phonon-electron scattering mechanisms and
show that phonon-electron scattering dominates at not too low
densities. We extract the ion thermal conductivity in ion liquid or
gas from literature.

Numerical values of the ion/phonon conductivity are approximated by
analytical expressions, valid for $T \ga 10^5\;{\rm K}$ and
$10^{5}~{\rm g~cm^{-3}}\la \rho \la 10^{14}~{\rm g~cm^{-3}}$.
Typical magnetic fields $B \sim 10^{12}$~G in neutron star envelopes
do not affect this conductivity although they strongly reduce the
electron thermal conductivity across the magnetic field. The ion
thermal conductivity remains much smaller than the electron
conductivity along the magnetic field. However, in the outer neutron
star envelope it can be larger than the electron conductivity across
the field, that is important for heat transport across magnetic
field lines in cooling neutron stars. The ion conductivity can
greatly reduce the anisotropy of heat conduction in outer envelopes
of magnetized neutron stars.

\end{abstract}



%
\section{Introduction}
\label{sect:introduct}

Neutron stars with strong magnetic fields $B\ga 10^{12}\;$G are
expected to have an anisotropic surface temperature distribution
(see, e.g., \citealt*{Geppert2004,Geppert2006,Perez2006}). It is
thought to result from the anisotropy of thermal conduction in a
neutron star envelope and manifests in periodic modulation of X-ray
emission observed from some spinning neutron stars (e.g.,
\citealt*{Burwitz2003,Haberl2007,Ho2007}). The thermal conductivity
$\kappa_\perp$ across the magnetic field can be much smaller than
the conductivity  $\kappa_\parallel$ along the field (see, e.g.,
\citealt*{YK,Potekhin_magn}). Both conductivities are crucial for
modeling of cooling magnetized neutron stars
(\citealt{Geppert2006,Perez2006}, \citealt*{Page2006}, and
references therein).

In a non-magnetized envelope, thermal energy is mainly transported
by electrons ($\kappa\approx \kappa_{\rm e}$;
\citealt{FI76,Potekhin_e_cond}). The thermal conductivity of  ions,
$\kappa_{\rm i}$, in which we include the ion conductivity of an ion
gas or liquid and the phonon conductivity of a crystalline ion
solid, is usually much smaller than $\kappa_{\rm e}$. However, a
strong enough magnetic field $B$ can easily suppress $\kappa_{{\rm
e}\perp}$, making $\kappa_{\rm i}$ the leading thermal conductivity
across the magnetic field (e.g., \citealt{Perez2006}).

In this paper we analyze different regimes of the ion thermal
conductivity in a neutron star envelope. Physical conditions in the
envelope are described in Sect.\ \ref{sect:plasma}. The thermal
conductivity in the magnetized envelope is outlined in Sect.\
\ref{sect-B}, where we give a sketch of anisotropic electron heat
conduction and nearly isotropic ion/phonon heat transport. The ion
conductivity due to ion-ion (phonon-phonon) scattering is studied in
Sect.\ \ref{sect-kappa_ii}. The ion conductivity owing to
ion-electron (phonon-electron) scattering is analyzed in Sect.\
\ref{sect:lattice-cond}. In Sect.\ \ref{sect:IonMag} we estimate
the values of magnetic fields which can strongly affect ion
heat transport. Section \ref{sect:discuss} contains a
discussion of our results.
Conclusion is presented in Sect.\
\ref{sect:conclusions}.

\section{Coulomb plasma in a neutron star envelope}
\label{sect:plasma}

\begin{figure}
    \centering
    \resizebox{3.2in}{!}{\includegraphics[bb=50 180 540 655]{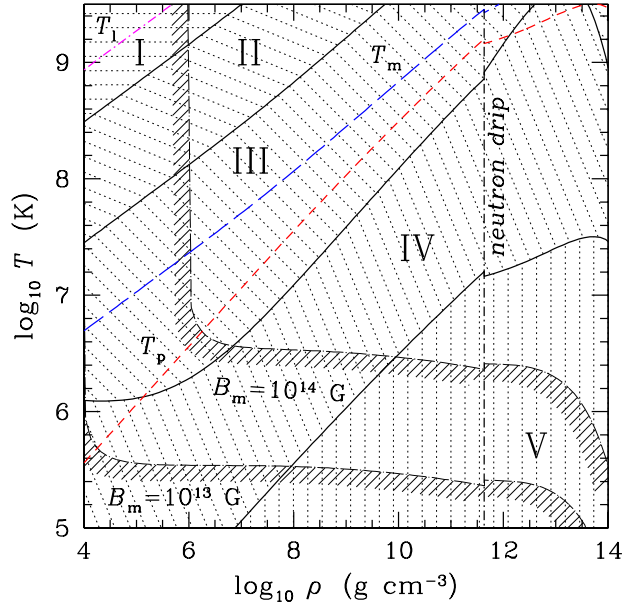}}
    \caption{(color online) Temperature-density diagram
        for the ground-state neutron star crust (smooth-composition model,
        \citealt{BOOK}); $T_\mathrm{l}$ is the temperature
    for the onset of Coulomb
        coupling of ions into ion liquid; $T_\mathrm{m}$ is the
        crystallization temperature of ions; $\Tp $ is the ion
        plasma temperature;
        the vertical line shows the neutron drip density. Shaded are regions
        I--V of different regimes (Table \ref{tab:regimes}) of the ion
        thermal conductivity. Also, we show the boundaries of
    the regions where ion heat transport can be strongly
    affected by the magnetic fields $B_\mathrm{m}=10^{13}$~G and
        $10^{14}$~G. See text for details.
    }
    \label{fig:diag}
\end{figure}

We consider a model of a neutron star envelope which combines
several models of ground state matter (\citealt{NV,Oyamatsu} and
others) and which is summarized by \citealt*{BOOK} (their Appendix
B). Typical densities and temperatures of practical interest are
indicated in Fig.\ \ref{fig:diag} and outlined below in this
section.  The shaded regions I--V are domains, where the ion thermal
conductivity has different character. These domains are summarized
in Table \ref{tab:regimes} and described further in Sects.\
\ref{sect-B}--\ref{sect:lattice-cond}. The last column of Table
\ref{tab:regimes} gives the leading ion (phonon) scattering
mechanism. In addition, in Fig.\ \ref{fig:diag} we show the
boundaries of the regions, where ion heat transport is strongly
affected by the magnetic fields $B_\mathrm{m}=10^{13}$~G and
$10^{14}$~G (the regions below and to the left of the lines).

At any density $\rho$ the plasma is assumed to consist of electrons
and single species of fully ionized atoms (bare atomic nuclei). In
the inner neutron star envelope, at densities higher the neutron
drip density $\rho_\mathrm{ ND}\approx 4.3 \times 10^{11}$
g~cm$^{-3}$ [shown by the vertical dot-dashed line in Fig.\
\ref{fig:diag}; \citealt{NV}] the matter contains also free
neutrons.

We describe the parameters of dense matter following \S 2.1 in
\citet{BOOK}.  The charge neutrality of the matter implies
\begin{equation}
    n_\mathrm{i}=n_\mathrm{e}/Z~,
\label{eq:n_i}
\end{equation}
where $Z$ is the charge number of the nuclei, while
$n_\mathrm{e}$ and $n_\mathrm{i}$ are the
electron and ion number densities, respectively.
The mass
density of the matter can be estimated as
\begin{equation}
    \rho\approx n_\mathrm{i}A' m_\mathrm{u}~,
\label{eq:rho_n_i_A}
\end{equation}
where $m_\mathrm{u}$ is the atomic mass unit and $A'$ is the number
of nucleons per one nucleus. For $\rho<\rho_\mathrm{ND}$, one has
$A'=A$, where $A$ is the number of nucleons confined in one nucleus.
For $\rho>\rho_\mathrm{ND}$, one has $A^\prime=A+A^{\prime\prime}$,
where $A^{\prime\prime}$ is the number of free (unbound) neutrons
per one nucleus. The ion mass is
\begin{equation}
  m_\mathrm{i}=A\,m_\mathrm{u}.
\end{equation}
%

\begin{table}
\caption[]{Ion conduction regimes I--V in Fig.\ \ref{fig:diag}.}
\label{tab:regimes}
\begin{center}
\begin{tabular}{clc}
\hline
Regime& State of ions & Leading \\ && scattering$^*$ \\
\hline
I & Weakly coupled ions &  ii   \\
II& Classical ion liquid & ii   \\
III & Low-$T$ liquid, high-$T$ crystal & ii (ph\,ph)  \\
IV  & Quantum ion crystal & ie (ph\,e) \\
V   & Very cold quantum crystal &  ie (ph\,e) and \\ && possibly others \\
\hline
\end{tabular}
\end{center}
\begin{center}
    $^*$Symbols i, e, and ph refer to ions, electrons,
and phonons, respectively.
\end{center}
\end{table}

First, we describe the properties of electrons (Sect.\ \ref{sect-e})
and ions (Sect.\ \ref{sect-i}) neglecting the effects of the magnetic
fields, and then (Sect.\ \ref{sect-B}) we outline the magnetic effects.

\subsection{Electrons}
\label{sect-e}

Under the conditions of study (Fig.\ \ref{fig:diag}), the electrons
are typically strongly degenerate. The electron Fermi momentum $\pF
$ and Fermi wavenumber $\kF $ are
\begin{equation}
   \pF = \hbar \kF =\hbar\,
   (3\pi^2 n_\mathrm{e})^{1/3}~.
\label{eq:p_F}
\end{equation}
%
The electron relativity
parameter can be written as
\begin{equation}
  \xr ={\pF \over m_\mathrm{e} c}
  \approx 1.009 \left({\rho_6 Z\over A^\prime}\right)^{1/3}~\;,
\label{xr}
\end{equation}
where $\rho_6\equiv  \rho/10^6~{\rm g~cm^{-3}}$.
The electron Fermi energy and the effective mass at the Fermi
surface take the form
\begin{equation}
   \epsilon_\mathrm{ F}=m_\mathrm{e} c^2(1+\xr ^2)^{1/2}\;,
   ~~ m_\mathrm{ e}^\ast
   =m_\mathrm{ e}\left(1+ \xr ^2\right)^{1/2}={\epsilon_\mathrm{ F}\over c^2}~,
\end{equation}
and the electron Fermi velocity reads
\begin{equation}
    \vF={\pF \over m^\ast_\mathrm{ e}} = c\, {\xr \over
    \left(1+\xr ^2\right)^{1/2}}~.
\end{equation}
The electron degeneracy temperature is
 \begin{equation}
 T_\mathrm{ F}={\epsilon _\mathrm{ F}-m_\mathrm{ e} c^2\over
 \kB }=5.93\times 10^{9}
 \left[\sqrt{1+\xr ^2}-1\right]\;{\rm K}\;,
 \end{equation}
where $\kB $ is the Boltzmann constant. The electrons are strongly
degenerate as long as $T\ll T_\mathrm{ F}$.

Finally, the electron Thomas-Fermi screening wavenumber (inverse
plasma screening length of Coulomb interaction due to polarizability of
strongly degenerate electrons) is
\begin{equation}
     \ktf=2\left(\alpha\;c\over \pi \vF\right)^{1/2}\kF ~,
\label{eq:DH.TF}
\end{equation}
where $\alpha$ is the fine structure constant.

\subsection {Ions}
\label{sect-i}

The properties of the classical ion plasma are determined by the
Coulomb coupling parameter
\begin{equation}
  \Gamma={Z^2 e^2\over \ai\kB T}~,
  \label{Gamma}
\end{equation}
where
\begin{equation}
\ai=\left(4\pi n_\mathrm{ i}/3\right)^{-1/3}~ \label{a_i}
\end{equation}
is the ion sphere radius. The plasma ions form a nearly ideal gas as
long as $\Gamma \ll 1$ which corresponds to $T \gg
T_\mathrm{l}=Z^2e^2/(\ai \kB)$.
At lower $T$ they transform smoothly
into a strongly coupled ion liquid.

It is also important to define the ion plasma frequency
\begin{equation}
  \op =\left(4\pi n_\mathrm{ i}Z^2\,e^2/m_\mathrm{ i}\right)^{1/2}~,
\end{equation}
and the associated ion plasma temperature
\begin{equation}
   \Tp ={\hbar \op \over \kB } =7.832\times
   10^6\;\left({\rho_6\over A^\prime}\; {Z^2\over A}\right)^{1/2}\;
   {\rm K}~,
\end{equation}
displayed in Fig.\ \ref{fig:diag}. Quantum effects in ion motion
become very important for $T\ll \Tp $.

The conditions of ion crystallization are described, e.g., in
\citet{BOOK}, \S 2.3.4. A classical Coulomb liquid of ions
crystallizes at the temperature
\begin{equation}
 T_\mathrm{ m}={Z^2 e^2\over \ai\kB \Gamma_\mathrm{
 m}}\approx 1.3\times 10^5\;Z^2\left({\rho_6\over
 A^\prime}\right)^{1/3}\; {175\over \Gamma_\mathrm{ m}}\;{\rm K}~,
 \end{equation}
where $\Gamma_\mathrm{ m}\approx 175$. Large zero-point vibrations
of ions at very high densities can reduce $T_\mathrm{ m}$ or even
prevent crystallization. Actually, this effect can be important only
for a plasma of light nuclei (hydrogen and helium) which we do not
consider here. Ions are usually assumed to form body-centered cubic
lattice, but, in fact, the face centered cubic lattice is also
possible (see, e.g., \citealt{Baiko2002}). We study phonon
conductivities for both lattice types and find, that they are nearly
identical. The same is true for other kinetic and thermodynamic
properties of such crystals (see, e.g.,
\citealt{Baiko1998,Baiko2001}).

\section{Thermal conductivity in a magnetized plasma}
\label{sect-B}

As outlined in Sect.\ \ref{sect:introduct}, we are interested in the
thermal conductivity of magnetized neutron star envelopes. A
magnetic field $\bm{B}$ makes thermal conduction anisotropic (e.g.,
\citealt{YK,Potekhin_magn}). The total thermal conductivity can be
written as
\begin{equation}
   \kappa=\kappa_\mathrm{e}+\kappa_\mathrm{i},
\label{kappa-total}
\end{equation}
where $\kappa_\mathrm{e}$ and $\kappa_\mathrm{i}$ are the
conductivities of electrons and ions, respectively. In the inner
envelope of a neutron star one should also add the thermal
conductivity of neutrons. However, it is largely unexplored and will
be neglected here. Generally, both conductivities,
$\kappa_\mathrm{e}$ and $\kappa_\mathrm{i}$, are anisotropic tensor
quantities. Nevertheless, a typical neutron star magnetic field $B
\sim 10^{12}-10^{13}$~G can strongly affect the electron thermal
conduction (see, e.g.,  \citealt{YK, Potekhin_magn}) but weakly
affects the ion thermal conduction (see Sec.\ \ref{sect:IonMag}).
Therefore, we will take into account the effect of the magnetic
field on $\kappa_\mathrm{e}$ but neglect its effect on
$\kappa_\mathrm{i}$. Let us emphasize that our aim is to study the
ion conductivity. We describe the electron conductivity here only
for comparison with the ion one.

\subsection{Thermal conductivity of electrons}
\label{sect:kappae}

An anisotropic electron conduction in a magnetic field $\bm{B}$ is
characterized by the thermal conductivity
$\kappa_\mathrm{e\parallel}$ along the magnetic field, by the
conductivity $\kappa_\mathrm{e\perp}$ across the field, and by the
Hall conductivity $\kappa_\mathrm{eH}$ (perpendicular to $\bm{B}$
and to the temperature gradient). The effects of the magnetic fields
on electron thermal conduction are twofold (see, e.g.,
\citealt{YK,VP}).

First, there are classical effects associated with electron Larmor
rotation about magnetic field lines. Their efficiency is
characterized by the electron magnetization parameter
$\omega_\mathrm{ g}\tau_\mathrm{e}$ (see e.g.\ \citealt{YK}), where
$\omega_\mathrm{ g}=eB/(m^\ast_\mathrm{ e} c)$ is the electron
gyrofrequency, and $\tau_\mathrm{ e}$ is the electron relaxation
time at $B=0$. In this classical approximation
\begin{equation}
  \kappa_\mathrm{e\parallel}=\kappa_\mathrm{e}^{(0)}
  ={\pi^2 \kB^2 T n_\mathrm{ e} \tau_\mathrm{ e} \over 3 m_\mathrm{ e}^*}~,
  \qquad \kappa_\mathrm{e\perp}=
  {\kappa_\mathrm{e}^{(0)}\over 1 +
  (\omega_\mathrm{g}\tau_\mathrm{ e})^2}~,
\label{kappae}
\end{equation}
and $\kappa_\mathrm{eH}=\kappa_\mathrm{e\perp}\,
\omega_\mathrm{g}\tau_{\rm e}$. If the electrons are strongly
magnetized ($\omega_\mathrm{g}\tau_\mathrm{ e}\gg 1$), their fast
Larmor rotation greatly reduces $\kappa_\mathrm{e\perp}$ and
$\kappa_\mathrm{eH}$. In this limit, $\kappa_\mathrm{eH}$ becomes a
non-dissipative quantity (independent of $\tau_\mathrm{e}$).

Second, electron transport can be modified by quantum effects
associated with the structure of electron Landau levels in the
$\bm{B}$-field. These effects are especially pronounced at
sufficiently low densities $\rho \lesssim \rho_{B\mathrm{e}}$, at
which degenerate electrons populate the only one (ground-state)
Landau level. The critical density $\rho_{B\mathrm{e}}$ can be
estimated as $\rho_{B\mathrm{e}}\equiv 7.045\times 10^{3}
(A/Z)(B/10^{12}{\rm G})^{3/2}\,{\rm g~cm^{-3}}$ (see, e.g.,
\citealt{Potekhin_magn}). At higher densities $\rho \gtrsim
\rho_{B\mathrm{e}}$, the electrons populate other Landau levels and,
when the density increases, all conductivity coefficients,
especially $\kappa_\mathrm{e\parallel}$ and
$\kappa_\mathrm{e\perp}$,
oscillate with growing $\rho$ around their classical values
(\ref{kappae}) in response to the population of new levels
\citep{Potekhin_magn}. These quantum oscillations are typically not
too strong and can be neglected here for our semi-quantitative
consideration of electron thermal transport. For simplicity, we will
restrict ourselves to high densities $\rho \gtrsim
\rho_{B\mathrm{e}}$ and neglect the quantum effects, but retain much
stronger classical effects.

In order to evaluate the electron conductivity coefficients in our
approximation we need the expression for $\tau_\mathrm{e}$. We will
calculate it including electron-ion scattering
(\citealt{Potekhin_e_cond,Gnedin}) [although neglecting
freezing of Umklapp processes 
discussed by \citealt{RaikhYakovlev,Gnedin}] 
and a recently revised contribution from electron-electron
scattering (\citealt{Shternin}). For this purpose, we will use the
Matthiessen rule $\tau_\mathrm{e}^{-1}=\tau_\mathrm{ei}^{-1}+
\tau_\mathrm{ee}^{-1}$ (see, e.g., \S 10 of Chapter 7 in
\citealt{Ziman}), where $\tau_\mathrm{ei}$ and $\tau_\mathrm{ee}$
are partial effective relaxation times (at $B=0$). Strictly
speaking, such an inclusion of electron-electron collisions into
transport coefficients (\ref{kappae}) in a magnetized plasma is
approximate. Similarly, at $T \lesssim \Tp $ electron-ion collisions
become essentially inelastic and Eq.~(\ref{kappae}) in a magnetized
plasma is quantitatively inaccurate even if electron-electron
collisions are neglected. However, we employ these approximations
for illustrating the importance of the ion thermal conductivity.

\subsection{Thermal conductivity of ions}
\label{sect:kappai}

As already mentioned above, we will neglect the
effects of magnetic fields on the thermal conductivity
of ions, and describe ion transport by one
coefficient of thermal conductivity, $\kappa_\mathrm{i}$
(for $B$=0). Estimates of magnetic field strengths
which make ion heat transport anisotropic are given in
Sect.\ \ref{sect:IonMag}. Their typical values are $\sim
10^{14}$~G, much higher than the magnetic fields
which induce a strong anisotropy of the electron transport.

The ion thermal conductivity can be presented as
\begin{equation}\label{kappa}
    \kappa_\mathrm{i}=\left( \kappa_\mathrm{ii}^{-1}+\kappa_\mathrm{ie}^{-1}
    \right)^{-1},
\label{ii+ie}
\end{equation}
where $\kappa_\mathrm{ii}$ and $\kappa_\mathrm{ie}$
are partial ion conductivities due to ion-ion and ion-electron
collisions, respectively. In fact, $\kappa_\mathrm{ie}$ appears to
be important only for crystallized ions (see Sec.\
\ref{sect:discuss}). There are several regimes of ion conduction
realized at different $T$ and $\rho$ (domains I--V in Fig.\
\ref{fig:diag} summarized in Table \ref{tab:regimes}). We analyze
them in subsequent sections.

Let us remark, that $\kappa_\mathrm{ie}$ should not be confused with
$\kappa_\mathrm{ei}$, the electron thermal conductivity due to
electron-ion (electron-phonon) scattering; $\kappa_\mathrm{ei}$ is
well known (e.g.\ \citealt{Gnedin}) and gives main contribution to
the electron transport.

\section{Ion conduction due to ion-ion collisions}
\label{sect-kappa_ii}

\subsection{Gaseous phase}
\label{sect-gas}

If $\Gamma \ll 1$ ($T \gg T_\mathrm{l}$ in Fig.\ \ref{fig:diag}),
the ions form a nearly ideal Boltzmann gas and the ion thermal
conductivity can be calculated from the standard Boltzmann equation
taking into account ion-ion and ion-electron Coulomb collisions. In
this case, ion-electron collisions are negligible and, according to
\cite{Braginski1963},
\begin{equation}
   \kappa_\mathrm{ii}^\mathrm{I} =3.9\,\frac{n_\mathrm{i}\,
   \tau_\mathrm{ii}^\mathrm{I}\,\kB^2   T}{m_\mathrm{i}}
   \approx 4\kappa_0\,\Gamma^{-5/2}/\Lambda_\mathrm{ii},
\label{cond-gas}
\end{equation}
where $\kappa_0=\kB\,\op\,n_\mathrm{ i}\, \ai ^2$ is a convenient
normalization constant,
\begin{equation}
   \tau_\mathrm{ii}^\mathrm{I}
   =\frac{3\,m_\mathrm{i}^{1/2}\,
   (\kB\,T)^{3/2}}{4\pi^{1/2}\,(Ze)^4\,n_\mathrm{i}\,\Lambda_\mathrm{ii}}
\label{tau-gas}
\end{equation}
is the effective relaxation time due to ion-ion collisions,
$\Lambda_\mathrm{ii}=\ln\left[1/(\sqrt{3}\Gamma^{3/2})\right]$ is
the appropriate Coulomb logarithm, and $\ai $ is the ion-sphere
radius given by Eq.\ (\ref{a_i}). In this regime, ion conduction is
realized through weak Coulomb collisions of almost free ions. In the
limit of $\Gamma \ll 1$ the Coulomb logarithm is sufficiently large
reflecting long-range nature of Coulomb interactions in the weakly
coupled plasma. For a moderate coupling ($\Gamma \sim 1$), one has
$\Lambda_\mathrm{ii} \sim 1$ because of the onset of strong ion
screening.

\subsection{Classical Coulomb liquid}
\label{sect-liquid}

%
Here we outline $\kappa_\mathrm{ii}$ in a classical strongly coupled
ion liquid ($1\lesssim \Gamma\lesssim \Gamma_\mathrm{m}$), where the
ions are no longer free but are mostly confined in Coulomb potential
wells. These results are expected to be especially suitable in
domain II in Fig.\ \ref{fig:diag}. In this case
$\kappa_\mathrm{ii}^\mathrm{II}$ can be calculated using the
molecular dynamics formalism. The physics of ion transport becomes
essentially different from the gaseous case (see, e.g.,
\citealt*{BV78,Pierleoni1987}). A nearly free ion motion in the gas
is replaced by oscillations in Coulomb potential wells, with
occasional hopping from one well to another (see, e.g.,
\citealt{Hopping}). Hopping transitions can contribute to the
effective thermal conductivity along with thermal energy transfers
via Coulomb interaction of ions vibrating in neighboring potential
wells. Such heat transport processes can be described in terms of
ion-ion scattering via emission and absorption of phonons in the ion
liquid; they can be studied by molecular dynamics technique (see,
e.g., \citealt{Ions_to_phons}). The resulting thermal conductivity
can be written as
\begin{equation}
  \kappa_\mathrm{ii}^\mathrm{II}
  =\kappa_\ast\,\kappa_0 \approx 0.4\, \kappa_0,
\label{cond-liquid}
\end{equation}
where $\kappa_\ast$ is a dimensionless, slowly variable function of
$\Gamma$. This function was calculated by \citet{BV78} and
\citet{Pierleoni1987} for $\Gamma\approx$1, 10, and 100, and the
results are in a good agreement (see Fig.\ \ref{fig:kappa_ii}). For
our simplified semiquantitative analysis, it is sufficient to set
this function equal to a typical value $\kappa_\ast=0.4$. It gives
us a temperature independent thermal conductivity
$\kappa_\mathrm{ii}^\mathrm{II}$ which combines smoothly with the
conductivity (\ref{cond-gas}) in the gaseous phase at $\Gamma \sim
1$. Expressing $\kappa_\mathrm{ii}^\mathrm{II}$ in the form of
familiar estimate $\kappa_\mathrm{ii} \sim n_\mathrm{i}
\tau_\mathrm{ii} \kB^2 T/m_\mathrm{i}$ [see Eq.\ (\ref{cond-gas})],
we obtain an estimate of the effective relaxation time,
$\tau_\mathrm{ii}^\mathrm{II} \sim \Gamma/\op$, which gives
$\tau_\mathrm{ii}^\mathrm{II} \sim \tau_\mathrm{ii}^\mathrm{I} \sim
\op^{-1}$ at $\Gamma \sim 1$.

\subsection{Coulomb crystal}
\label{sect-crystal}

Now let us study $\kappa_\mathrm{ii}$ in domains III--V in Fig.\
\ref{fig:diag}, which mainly refer to crystalline matter. In this case,
ion transport can be described using the formalism of phonons
(elementary
excitations of
the crystalline lattice; \citealt{Ziman}, Chapter 1) and
$\kappa_\mathrm{ii}$ is determined by phonon-phonon scattering
(absorptions and emissions of phonons, ph, which can be
schematically presented as $\mathrm{ph+ph\to ph}$ and
 $\mathrm{ph \to ph+ph}$).
For the parameters of dense matter we are interested
in (Fig.\ \ref{fig:diag}), the approximation of almost uniform
electron background holds (e.g., \citealt*{Potekhin97}). Then the
phonon spectrum and anharmonic terms (which determine phonon-phonon
scattering) of Coulomb crystals can be computed with high precision
(e.g., \citealt{CohenKeffer55,Dubin90}). Accordingly, one can
accurately calculate $\kappa_\mathrm{ii}$ but this calculation is
complicated and we restrict ourselves by semi-quantitative
estimates. Note that phonon scattering in the crystal is similar to
that in the ion liquid although the properties of phonons in the
liquid and the crystal can be somewhat different (e.g., \S 71 in
\citealt{Stati1}, and \S 24 in \citealt{Stati2}).

For a classical Coulomb crystal ($\Tp /3 \lesssim T \lesssim
T_\mathrm{m}$) we employ a simple estimate by \cite{Ziman} [his Eq.\
(8.2.16) in Chapter 8]
\begin{equation}
\label{kappa_phph_Ziman}
        \kappa_\mathrm{ii}^{\rm cl\, latt}
    \approx \frac{m_\mathrm{ i}n_\mathrm{ i}\, \cs ^3\,\ai }
        {\gamma^2\,T}\approx \frac{4\,m_\mathrm{ i} n_\mathrm{ i}\,
    \op ^3}
        {27\,T\,\qBZ ^4}.
\label{cl-latt}
\end{equation}
Here,  $\gamma= \partial \ln \Tp /\partial\ln V =0.5$ is the Gr\"
uneisen constant and we assume that the typical phase velocity of
phonons is $\cs \approx \op /(3\, \qBZ )$ [we replace the Brillouin
zone by a sphere with the radius $\qBZ =(6\pi^2\,n_\mathrm{
i})^{1/3}$]. \cite{Perez2006} estimated $\kappa_\mathrm{ii}^{\rm
cl\,latt}$ from the same expression but using $\gamma=2$,
characteristic for terrestrial solids (where the electron background
is essentially non-uniform). Introducing our
conductivity normalization constant
 $\kappa_0$, from Eq.\
(\ref{cl-latt}) we obtain
\begin{equation}
   \kappa_\mathrm{ii}^{\rm cl\,latt}=
   {8 \over 243 \pi} \,\left(6 \over \pi\right)^{1/3}\Gamma \kappa_0
   \approx {\Gamma \over 77}\,\kappa_0.
\label{cl-latt1}
\end{equation}

The thermal conductivity $\kappa_\mathrm{ii}$ in a crystal is solely
determined by Umklapp phonon-phonon scattering (the Peierls theorem;
e.g., Chapter 8 in \citealt{Ziman}), in which the sum of two (either
initial or final) phonon wavevectors
goes outside the first Brillouin zone. In addition, there are so
called normal scattering processes, where the sum of wavevectors
remains in the first Brillouin zone. With decreasing $T$ (or
increasing $\rho$) the Umklapp processes become frozen out because
they involve phonons with $\omega\sim\op $ (whereas the amount of
such phonons at $T\ll \Tp $ -- in the quantum, low-temperature limit
-- is exponentially small). The freezing strongly enhances
$\kappa_\mathrm{ii}$.

Calculations of $\kappa_\mathrm{ii}$ at $T \ll \Tp $ are difficult;
some estimates of $\kappa_\mathrm{ii}$ under different assumptions
are presented in Chapter 8 of \citet{Ziman}. However, for a neutron
star crust it is sufficient to use a simpler estimate, consisting in
multiplying the conductivity (\ref{cl-latt1}) for a classical
crystal by an appropriate enhancement factor,
\begin{equation}
\label{kappa_phph}
    \kappa_\mathrm{ii}^{\mathrm{III}\textrm{-}\mathrm{V}}\approx
    \kappa^{\rm cl\,latt}_\mathrm{ii}
    \,\exp\left(\beta \Tp/T\right),
\end{equation}
where  $\beta \Tp\sim \Tp/3$ corresponds to the estimation of the
lowest frequency of phonons participating in Umklapp processes.
We will use this expression everywhere in domains III--V. The
enhancement factor $\exp\left(\beta\Tp/T\right)$ can be inaccurate
at $T\ll \Tp$ but the main contribution into the total ion
conductivity $\kappa_\mathrm{i}$ at these low $T$ comes from
ion-electron scattering (Sect.\ \ref{sect:lattice-cond}), so that
the inaccuracy does not strongly affect the total ion conductivity
$\kappa_\mathrm{i}$.

Note that the phonon thermal conductivity can be presented in the
form (\citealt{Ziman}, \S1 in Chapter 7)
\begin{equation} \label{kappa_ie_stand}
  \kappa_\mathrm{i}
  \approx \frac{1}{3}\,\kB\,C_\mathrm{i} n_\mathrm{i} \cs
  L_\mathrm{ph},
\end{equation}
where $C_\mathrm{i}$ is the phonon (dimensionless) heat capacity per
one ion (see, e.g., \citealt{Baiko2001}); $\cs \approx \op /(3\qBZ)$
is a typical phonon group velocity, and $L_\mathrm{ph}$ is an
effective phonon mean free path. Equation (\ref{kappa_phph}) gives
\begin{equation}
 L_\mathrm{ph}\approx 20\,\ai
    \frac{\Gamma}{\Gamma_\mathrm m}\,\frac{3}{C_\mathrm{i}}\,
    \exp\left(\beta \Tp/T\right).
\end{equation}
We remark that $C_\mathrm{i}\approx 3$ for $T\gtrsim \Tp /3$ and
$C_\mathrm{i}\sim (T/\Tp )^3$ for $T\ll \Tp /3$ (see, e.g.,
\citealt{Baiko2001}).  Let us emphasize that the effective mean free
path $L_\mathrm{ph}$ should be calculated neglecting normal
phonon-phonon scattering processes which do not affect phonon heat
transport (\citealt{Ziman}, Chapter 8).

At $T \lesssim \Tp $, an important contribution into
$\kappa_\mathrm{i}$ can result from phonon-impurity scattering. At
very low temperatures, phonon scattering by crystal boundaries can
become significant (the neutron star crust can be a polycrystal made
of small monocrystals).  However, in view of our ignorance
concerning the actual nature and distribution of impurities and
polycrystal structures in the crust, these mechanisms will not be
considered in the present paper. In the high-temperature limit
($T\gtrsim \Tp /3$), we obtain a not too large mean free path
$L_\mathrm{ph}\sim 20
\ai $.
It indicates that the presence of not very abundant impurities (a
few percent by number) or bulky monocrystals will not affect the ion
thermal conductivity. In the low-temperature quantum limit ($T\ll
\Tp /3$) the phonon mean free path is restricted by electron
scattering (Sec. \ref{sect:lattice-cond}) and can be estimated as
$L\sim
200 \ai $
[Sec.\
\ref{sect:results}, Eq.\ (\ref{Fit})]. In this case
heat is mainly conducted by long wavelength phonons (Sec.\
\ref{sect:U-low-T}), whose scattering by impurities is inefficient
(\citealt{Ziman}, Chapter 8, \S 3). This means that the impurities
are not very important at low temperatures as well.

\subsection{Interpolation expression for $\kappa_\mathrm{ii}$}
\label{Sec:interpol}

\begin{figure}
    \centering
    \resizebox{3.2in}{!}{\includegraphics[]{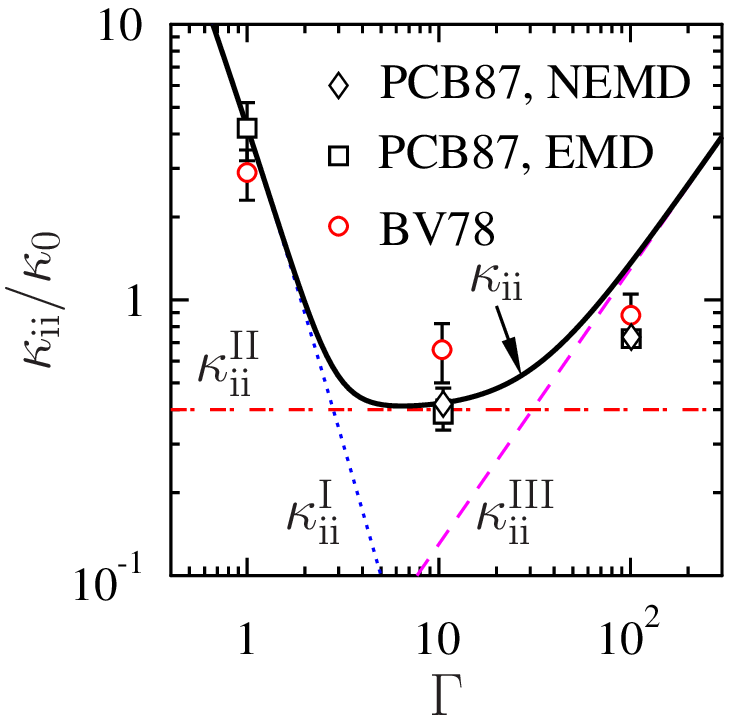}}
    \caption{(color online) Dimensionless ion-ion thermal
    conductivity $\kappa_\mathrm{ii}/\kappa_0$
    for a classical ion plasma ($T\gg \Tp $)
    as a function of $\Gamma$. Circles show results of
    equilibrium molecular dynamics (EMD) simulations of \citealt{BV78} (BV 78),
    squares
    and diamonds present equilibrium  and nonequilibrium
    molecular dynamics results of
    \citet{Pierleoni1987} (PCB87 EMD and NEMD), respectively.
    The solid line is our interpolation, Eq.\ (\ref{interpol}). Dotted,
    dot-dashed and dashed lines are the conductivities in domains
    I--III, given by Eqs.\ (\ref{cond-gas}), (\ref{cond-liquid}) and
    (\ref{cl-latt1}), respectively.
    }
    \label{fig:kappa_ii}
\end{figure}

To facilitate the implication of the above results we can
suggest the following interpolation formula for
$\kappa_\mathrm{ii}$ in all domains I--V
in Fig.\ \ref{fig:diag},
\begin{eqnarray}
 \kappa_\mathrm{ii}
 &=&
 \left\{\left(\kappa_\mathrm{ii}^\mathrm{I}\right)^2
 +\left(\kappa_\mathrm{ii}^\mathrm{II}\right)^2
 +\left(\kappa_\mathrm{ii}^{\mathrm{III}\textrm{-}\mathrm{V}}
  \right)^2\right\}^{1/2}
\nonumber \\
 &=&
 \kappa_0\,
 \left\{
      \Gamma^{-5}/\ln^2\left[2+1/(\sqrt{3}\Gamma^{3/2})\right]
  \right.
 \nonumber \\
   &+&  \left.
      \kappa_{\ast}^2
      +
      \left(\Gamma/77\right)^2
      \,\exp\left(2\beta \Tp/T\right)
      \right\}^{1/2},
\label{interpol}
\end{eqnarray}
with $\kappa_{\ast}\approx 0.4$ and $\beta \approx 1/3$. Here, we
assume a smooth temperature variation of $\kappa_\mathrm{ii}$,
without any jump or break at the melting point $T=T_\mathrm{m}$. Our
assumption is made in analogy with the electron thermal conductivity
which does not show any peculiarity at $T=T_\mathrm{m}$
\citep{Baiko1998} because of the importance of multi-phonon
electron-phonon scattering processes in crystalline lattice at $T <
T_\mathrm{m}$ and because of incipient ion-ion correlations in ion
liquid at $T>T_\mathrm{m}$. Another argument in favor of
a relatively smooth
behavior of the conductivity near the melting point is given
by the presence of shear modes in a strongly coupled Coulomb
liquid
\citep{Schmidt1997}, that is typical for a crystal rather than for
a liquid.

Figure \ref{fig:kappa_ii} shows the dependence of a dimensionless
thermal conductivity $\kappa_\mathrm{ii}/\kappa_0$
in a classical ion plasma on the Coulomb coupling parameter $\Gamma$.
Circles are the results of \citet{BV78}
(BV78). Squares and diamonds are calculations of \citet{Pierleoni1987}
carried out using equilibrium molecular dynamics and
nonequilibrium molecular dynamics techniques (PCB87
EMD and PCB87 NEMD), respectively. Dotted, dot-dashed and dashed
lines present the thermal conductivities
calculated, respectively, from Eqs.\
(\ref{cond-gas}), (\ref{cond-liquid}) and (\ref{cl-latt1}),
which are valid in domains I, II and III.
Our interpolation (\ref{interpol}), plotted by the solid line,
is in good agreement
with the results of \citet{BV78} and \citet{Pierleoni1987}.
In particular, the interpolation predicts a minimum of
$\kappa_\mathrm{ii}/\kappa_0$ at $\Gamma\sim 7$, which agrees
with the estimate of \citet{BV78} that the minimum takes place
in the vicinity of $\Gamma=10$. The largest, but still acceptable
deviation of the interpolated $\kappa_\mathrm{ii}/\kappa_0$
from the results of \citet{BV78} and \cite{Pierleoni1987}
occurs at $\Gamma\sim 100$. It can be explained by a significant
contribution of multi-phonon processes [neglected
in Eq.\ (\ref{cl-latt1})], as well as by an inaccuracy of
Eq.\ (\ref{cl-latt1}), which is just an order-of-magnitude
estimate of the thermal conductivity of ions in the crystalline
phase that we use to determine the thermal conductivity at
$\Gamma= 100$ in our interpolation.

\section{Phonon conduction due to phonon-electron scattering}
\label{sect:lattice-cond}

Now let us discuss $\kappa_\mathrm{ie}$, which is the partial ion
thermal conductivity due to ion-electron scattering in Eq.\
(\ref{ii+ie}). This process was not considered by \cite{Perez2006}.
As already mentioned above, this mechanism appears to be
important only for the crystalline lattice at
$T\la T_\mathrm{m}$, and becomes negligible at higher $T$
as compared to ion-ion scattering.

\subsection{General formalism}
\label{sect:formalism}

We will mainly focus on the case in which $T \lesssim T_\mathrm{m}$
and the phonon formalism is appropriate, so that
$\kappa_\mathrm{ie}$ can be treated as the phonon thermal
conductivity produced due to phonon-electron scattering (due to
absorptions and emissions of phonons by electrons, ph+e$\to$e and
e$\to$ph+e). We expect that the formalism described below is valid
for calculating $\kappa_\mathrm{ie}$ in temperature-density domains
III--V displayed in Fig.\ \ref{fig:diag} and listed in Table
\ref{tab:regimes}. The conductivity $\kappa_\mathrm{ie}$ can be
calculated using the variational method, described, for instance, in
Chapter 7 of
\citet{Ziman}. The variational expression reads
\begin{equation}
    \kappa_\mathrm{ie}= {1 \over \dot{S}_\mathrm{coll}}\;
    \left|\sum_s \int \bm{v}_{\nu} \Phi_{\nu}
            \frac{\partial n_{\nu}}{ \partial T}\,{\rm d}\bm{Q}
            \right|^2.
\label{kappa_univ}
\end{equation}
In this case, $\nu\equiv ({\bm Q},\,s)$ enumerates phonon modes,
$\bm{Q}$ is a phonon wavevector, and $s$ is a polarization index; a
phonon frequency will be denoted by $\omega_\nu$. Furthermore,
$n_{\nu}=1/[\exp(\hbar\omega_\nu/\kB  T)-1]$ is the phonon
equilibrium occupation number, and
 ${\bm v}_{\nu}=\partial \omega_\nu/\partial {\bm Q}$ is
the phonon group velocity;  $\dot{S}_\mathrm{coll}$ is the
entropy generation rate
in phonon-electron scattering. A variational function $\Phi_{\nu}$
describes a weak deviation of the phonon distribution function from
the equilibrium distribution $n_{\nu}$. The variational estimate of
$\kappa_\mathrm{ie}$ given by Eq.~(\ref{kappa_univ}) reaches maximum
for the exact solution of the phonon transport equation. For
phonon-electron scattering, the entropy generation rate can be
expressed as (\citealt{Ziman}, Chapter 8, \S 9)
\begin{eqnarray}
\label{Sdot_univ}
    \dot{S}_\mathrm{coll}&=&\frac{1}{\kB \,T^2} 
    \int 
        \Phi_{\bm{Q}}^2
    \mathcal{P}_{{\bm Q},{\bm k}}^{\bm{ k'}}
    {\rm d} {\bm Q}\, {\rm d} {\bm k}\,  {\rm d} \bm{k'}.
\end{eqnarray}
%
For the sake of simplicity, polarization indices are omitted here. Wavevectors
$\bm{k}$ and ${\bm k}^\prime$ refer to electrons.
In calculating $\kappa_\mathrm{ie}$ the electron
distributions are assumed to be equilibrium ones.
Finally,
$\mathcal{P}_{{\bm Q},{\bm k}}^{{\bm k}^\prime}$ is
the differential transition probability, calculated in equilibrium.

The derivation of $\kappa_\mathrm{ie}$ is simplified
(\citealt{Ziman}, Chapter 8, \S 9)
by the equality of the squared matrix elements for the phonon
absorption and emission processes. Employing the simplest suitable
variational function $\Phi_\nu=\bm{Q}\cdot\bm{u}$ ($\bm{u}$ being
the unit vector along the temperature gradient),
one gets
%
\begin{equation}
\label{kappa_phe0}
    \kappa_\mathrm{ie}=\left(\frac{\kB \,C_\mathrm{i}}
    {3\,Z\,e}\right)^2\,T\,\widetilde{\sigma},
\end{equation}
where
%
\begin{equation}\label{S_col_e}
    \widetilde{\sigma}=\frac{e^2\,n_\mathrm{e}\widetilde\tau_\sigma}{m^\ast_e},
    \quad
    \widetilde\tau_\sigma=
    \frac{\pF ^2\vF }{4\pi\,Z^2e^2\,n_\mathrm{i}\widetilde\Lambda_\mathrm{e}},
\end{equation}
are auxiliary quantities,
which are formally similar (but not identical) to the electron electric
conductivity and the electron effective relaxation time
due to electron-photon scattering,
and
\begin{eqnarray}
\label{Lambda}
\widetilde\Lambda_\mathrm{e}
 &=&
 \frac{2\,\pF ^2}{m_\mathrm{i} \kB T}
    \,\int\frac{{\rm d}\Omega_{\bm{k}}}{4\pi}
    \,\int \frac{{\rm d}\Omega_{\bm{k'}}}{4\pi}
    \left|\phi(q)\right|^2\,
    \nonumber \\
   &\times&
    \left[1-\frac{\vF ^2}{c^2}\,
    \left(\frac{q}{2\kF }\right)^2\right]\, Q^2\, \exp[-W(q)]
     \nonumber \\
   &\times&
       \sum_{s} ({\bm q}\cdot\bm{e}_\nu)^2
    \,   \frac{\exp(-z_\nu)}{\left[1-\exp(-z_\nu)\right]^2}.
\end{eqnarray}
is formally similar to the Coulomb logarithm for the electron
transport [although it differs from the real Coulomb logarithm that
contains $q^2$ instead of $Q^2$; see, e.g., \citealt{Baiko1995}].
 Since $\widetilde{\sigma}$ and $\widetilde\tau_\sigma$ are
 not real electric conductivity and relaxation time,
 $\kappa_\mathrm{ie}$ and $\widetilde{\sigma}$ do not obey
 the  Wiedemann-Franz law. 
The integration in (\ref{Lambda}) is done over positions of
both electrons on the Fermi surface, with wavevectors
$\bm{k}$ and $\bm{k'}$ ($\mathrm{d}\Omega_{\bm{k}}$ and
$\mathrm{d}\Omega_{\bm{k'}}$ are respective solid angle
elements). Furthermore, $\bm{q}=\bm{k}-\bm{k'}$ is an
electron momentum transfer in a scattering event, and
$\bm{Q}$ is the phonon wavevector that is equal to $\bm{q}$
reduced to the first Brillouin zone (see Chapter 1 of
\citealt{Ziman} for more details). Specifically,
$\bm{Q}=\bm{q}-\bm{g}$, where $\bm{g}$ is a reciprocal
lattice vector which realizes such a reduction (absorbs a
momentum excess). Let us remind that the processes with
$\bm{g} \neq \bm{0}$ belong to Umklapp phonon-electron
scattering processes, while the processes with
$\bm{g}=\bm{0}$ are normal phonon-electron processes. The
quantity $\left|\phi(q)\right|\approx F(q)/(q^2+\ktf^2)$ in
Eq.~(\ref{Lambda}) is a
Fourier transform
 of the screened Coulomb
electron-ion interaction; $\ktf$ is
 the Thomas-Fermi screening wavenumber
 given by Eq.\ (\ref{eq:DH.TF});
$F(q)$ is the nucleus form factor which accounts for the proton
charge distribution within the atomic nucleus; $W(q)$ is the doubled
Debye-Waller factor (see, e.g., \citealt{Baiko1995}). In addition,
we have defined $z_\nu=\hbar\omega_s(\bm{Q})/\kB T$ and the phonon
polarization vector $\bm{e}_\nu=\bm{e}_s({\bm  Q})$.

We have performed extensive computations of
$\widetilde{\Lambda}_\mathrm{e}$ and $\kappa_\mathrm{ie}$ from Eqs.\
(\ref{kappa_phe0})--(\ref{Lambda}) by the Monte Carlo technique. In
these calculations, the positions of electrons ($\bm{k}$ and
$\bm{k'}$) on the Fermi surface are randomly selected at every step
in long Monte Carlo runs. For every selection, we calculate
$\bm{q}$, a respective inverse lattice vector $\bm{g}$, and a phonon
wavevector $\bm{Q}$. The phonon eigenfrequencies $\omega_s(\bm{Q})$
and polarization vectors $\bm{e}_s(\bm{Q})$ have been determined
then for a Coulomb crystal of ions immersed in the rigid electron
background (see, e.g., \citealt{Baiko2001}).
Small
(but finite) widths of energy gaps in electron energy spectrum at
the intersections of the electron Fermi surface with boundaries of
Brillouin zones have been neglected. We have calculated
$\kappa_\mathrm{ie}$ for body-centered cubic and face-centered cubic
crystals, and the results turn out to be the same. The same
conclusions have been reached by \citet{Baiko1998} with regard to
the electron transport coefficients determined by electron-phonon
scattering.

When the temperature decreases, the number of phonons, which can
efficiently participate in phonon-electron scattering, becomes
smaller. They can only be the phonons with small frequencies
[$\hbar\omega_s(\bm{Q})\sim \kB T$] and, hence, with small
wavevectors $\bm{Q}$ near the center of the Brillouin zone. The
regions on the Fermi surface, which mainly contribute to the
integral (\ref{Lambda}), become narrower, and the accuracy of our
direct Monte Carlo calculations gets lower. At sufficiently low $T$
direct Monte Carlo calculation is inefficient and we replace it with
a semi-analytic consideration (Sect.\ \ref{sect:U-low-T}). In Fig.\
\ref{fig:diag}, the $T-\rho$ domains, where direct Monte Carlo
calculations are effective, are labeled as III and IV, while domain
V will be described semi-analytically. We combine Monte Carlo and
semi-analytic results and produce a useful practical fit expression
in Sect.\ \ref{sect:results}.

\subsection{Umklapp processes at low temperatures}
\label{sect:U-low-T}

Let us analyze phonon-electron  Umklapp processes (${\bm g}\ne
\bm{0}$) at low temperatures (domain V in Fig.\ \ref{fig:diag}). In
this case it is sufficient to consider small phonon wavenumbers
$Q\ll \qBZ $, and we can rewrite Eq.~(\ref{Lambda}) in the form
\begin{eqnarray}
\widetilde\Lambda^{\rm U}_\mathrm{e}
 &=&
\frac{\pF ^2}{8\,\pi^2\,m_\mathrm{i} \kB T}\sum_{g<2\kF }
     \frac{F^2(g)}{g^4}\,\left[1-\frac{\vF ^2}{c^2}\,
     \left(\frac{g}{2\kF }\right)^2\right]\cr\cr
    &\times&\exp[-W(q)] I_{\bm g},
\label{LambdaI} \\
     I_{\bm g}&=&\int{\rm d}\Omega_{\bm{k}}
    \,{\rm d}\Omega_{\bm{k'}}
    \,Q^2 \,
    \sum_{s} ({\bm g}\cdot{\bm e}_\nu)^2
    \,\frac{\exp(-z_\nu)}{\left[1-\exp(-z_\nu)\right]^2},
\label {LambSum}
\end{eqnarray}
where the sum is over all non-zero reciprocal lattice
vectors inside the Fermi sphere ($g<2\kF $). 
Using symmetry properties,
one can show that $(\bm{g}\cdot\bm{e}_\nu)^2$ in the integrand can be
replaced by $g^2/3$.

At low temperatures we are interested in, the main contribution into
$I_{\bm g}$ comes from those positions of $\bm{k}$ and $\bm{k'}$ on
the Fermi sphere, for which $\bm{k}-\bm{k'} \approx \bm{g}$. This
case can be studied semianalytically following \citet{PT97} who
considered analogous problem in the neutrino emission due to
electron-nucleus collisions. Let us introduce a coordinate system
whose $z$ axis is directed along $\bm{g}$. Let us further define
$\bm{k}$ by its polar and azimuthal angles $\vartheta$ and
$\varphi$, and $\bm{k'}$ by its polar and azimuthal angles
$\vartheta'$ and $\varphi'$, and introduce $\mu=\cos \vartheta$,
$\mu'=\cos \vartheta'$. The allowed positions of $\bm{k}$ should
concentrate
on
 a ring ($0 \leq \varphi < 2 \pi$)
with $\mu \approx \mu_0=g/(2\kF )$; while the respective positions
of $\bm{k'}$ should concentrate
on
a complementary ring ($\varphi' \approx \varphi\equiv \varphi_0$,
$\vartheta' \approx \pi - \vartheta$). Let $\bm{k}_0$ be a position
of $\bm{k}$ exactly on the ring, and
$\bm{k'}_0=\bm{k}_0-\bm{g}$
be a respective ($\varphi'=\varphi$)
position of $\bm{k'}$ on the complementary ring. Small variations of
$\bm{k}$ and $\bm{k'}$ around these positions can be defined by
$\dm= \mu-\mu_0$, $\dmp= \mu^\prime-\mu'_0$, $\dvp=
\varphi^\prime-\varphi_0$,
$\Delta{\bm k}=\bm{k}-\bm{k}_0$, and
 $\Delta \bm{k'}=\bm{k'}-\bm{k'}_0$.
Using cylindrical coordinates, we get then
\begin{eqnarray}
 \Delta k_{||}\approx \kF \,\dm,\quad
 \Delta k_{\bot}\approx -\frac{\mu_0}{\sqrt{1-\mu_0^2}}\,\kF \,\dm,
 &&
 \Delta k_{\varphi}=0,
 \cr\cr
 \Delta k_{||}^\prime\approx \kF \,\dmp,\quad
 \Delta k_{\bot}^\prime\approx
 \frac{\mu_0}{\sqrt{1-\mu_0^2}}\,\kF \,\dmp,&&
 \cr\cr
 \Delta k_{\varphi}^\prime\approx\sqrt{1-\mu_0^2}\,\kF \dvp.&&
\end{eqnarray}
Because the phonon wavevector is
$\bm{ Q}=\Delta {\bm k}-\Delta {\bm k}^\prime$, we obtain
\begin{eqnarray}
Q_{||}&\approx& \kF \,\left(\dmp+\dm\right),\quad\cr\cr
Q_{\bot}&\approx& \frac{\mu_0}{\sqrt{1-\mu_0^2}}\,\kF
\,\left(\dmp-\dm\right),\cr\cr
 Q_{\varphi}&\approx&
-\sqrt{1-\mu_0^2}\,\kF \,\dvp.
\end{eqnarray}
The Jacobian of the transformation from variables
$(\dm,\,\dmp,\, \dvp)$ to $(Q_{||},\, Q_\bot,\, Q_{\varphi})$ is
 $1/{2\,\mu_0\kF ^3}$. It is then easy to pass
 from coordinates   $(Q_{||},\, Q_\bot,\, Q_{\varphi})$
to spherical coordinates of the vector $\bm{Q}$.
Because only
small-$Q$ phonons
are available in the cold crystal,
we can safely extend the integration over $Q$ to infinity.
As a result, we obtain
\begin{eqnarray}
\label{I_q}
 I_{\bm g}&=& \frac{g}{3\,\kF ^2}\, \int_{0}^{2 \pi} {\rm d} \varphi \sum_{s}
 \int {\rm d}\Omega_{\bm Q}
 \int_0^\infty {\rm d} Q\,Q^4 \, \cr\cr
    &~& \times \,\frac{\exp(-z_\nu)}{\left[1-\exp(-z_\nu)\right]^2}.
\end{eqnarray}
At  $Q\ll \qBZ $ in the Coulomb crystal we have two nearly acoustic
phonon modes ($z_{1,\,2}\approx \hbar\,c_{1,\,2}(\hat{\bm Q})\,Q/\kB
T$, $\hat{\bm Q}$ being a unit vector along $\bm{Q}$) and one nearly
optical mode ($z_3\approx \hbar \op /\kB T\gg 1$) that does not
contribute to $I_{\bm g}$. The
 integration over ${\rm d}Q$ for acoustic modes gives
\begin{eqnarray}
 \int_0^\infty Q^4\, \frac{\exp(-z_{1,\, 2})}
  {\left[1-\exp(-z_{1,\, 2})\right]^2}\,{\rm d}Q
 &=&4!\,\zeta(4)\,\left[\frac{\kB \,T}{\hbar\,
 c_{1,\, 2}(\hat{\bm Q})}\right]^5=\cr\cr
 \quad=\frac{4\pi^4}{15}\,
 \left[\frac{\kB \,T}{\hbar\,c_{1,\, 2}(\hat{\bm
 Q})}\right]^5,&&
\end{eqnarray}
and the integral over $\varphi$ gives $2\pi$. Therefore,
\begin{eqnarray}
 I_{\bm g} &=& \frac{8\pi^5}{45}\,\left(\frac{\kB \,T}{\hbar}\right)^5
 \frac{g}{\kF ^2}
  \sum_{s=1,\,2}\int {\rm d}\Omega_{\bm Q}
 \,c_s^{-5}\,=\cr\cr
 &=&\frac{64\pi^6}{45}\,\left(\frac{\kB \,T}{\hbar\,\overline{\cs}}\right)^5\,\frac{g}{\kF ^2}
 ,
\end{eqnarray}
where
\begin{equation}
\overline{c}_\mathrm{s}\equiv  \left({1\over 8\pi}\,\sum_{s}\int
{\rm d}\Omega_{\bm Q}\,c_s^{-5}\right)^{-1/5}
\end{equation}
is an average phonon group speed that can be written as
$\overline{\cs}={\cal A}\,\op /\qBZ $. The calculation yields
 ${\cal A}\approx 0.36$
for both bcc and fcc Coulomb crystals.
Finally, we have
\begin{eqnarray}
\label{Lambda-low-T}
 \widetilde\Lambda^\mathrm{U}_\mathrm{e}&=&
 \frac{\pi^4}{45}\,\frac{\hbar\,\qBZ ^2}{m_\mathrm{i}\,\op }
 \,\left(\frac{\qBZ }{\kF }\right)^3\,
 {1 \over {\cal A}^{5}}\,\left(T \over \Tp \right)^4\sum_{g<2\kF }
     F^2(g)\times
 \cr\cr
     &\times&\left(\frac{2 \kF }{g}\right)^3
     \left[1-\frac{\vF ^2}{c^2}\,\left(\frac{g}{2\kF }\right)^2\right]
    \, \exp[-W(q)].
\end{eqnarray}
As in the case of electron transport (see, e.g.,
\citealt{Potekhin_e_cond}), normal phonon-electron processes give
small contribution
into scattering rate
 and can be neglected. The same is
true for higher temperatures.

Notice in passing, that in the same manner we can obtain an
asymptotic form of the Coulomb logarithm for the electron electrical
conductivity due to electron-phonon scattering,
\begin{eqnarray}
\label{Lambdaeph} {\Lambda^{\sigma}_\mathrm{ U}}&=&
 \frac{\pi^2}{9}\,\frac{\hbar\,\qBZ ^2}{m_\mathrm{i}\,\op }
 \,
 \frac{\qBZ }{\kF }
 \,
 {1 \over {\cal A'}^{3}}\,\left(T \over \Tp \right)^2
 \sum_{g<2\kF }
     F^2(g)\,
     \frac{2 \kF }{g}
     \cr\cr
&\times&     \,\left[1-\frac{\vF
^2}{c^2}\,\left(\frac{g}{2\kF }\right)^2\right]
    \, \exp[-W(q)].
\end{eqnarray}
However, in this case an effective phonon group velocity is
$\overline{\cs'}={\cal A}'\,\op /\qBZ $ with ${\cal A}'\approx 0.40$
for both bcc and fcc lattices. The difference in the values of
${\cal A}$ and ${\cal A}'$ results from averaging different powers
of $\cs (\hat{\bm Q})$. Our asymptotic expression for
${\Lambda^{\sigma}_\mathrm{U}}$ agrees with previous calculations
(e.g., \citealt{Gnedin}).

\subsection{Interpolation expression}
\label{sect:results}

In order to approximate the thermal conductivity $\kappa_\mathrm{ie}$
in domains III--V we present it in the familiar form
\begin{equation} \label{kappa_ie_approx}
\kappa_\mathrm{ie}
  =\frac{1}{3}\,\kB\,C_\mathrm{i} n_\mathrm{i} \cs
  L_\mathrm{ph}^\mathrm{ie},
\end{equation}
with $\cs =\op /(3\qBZ )$. We have extracted the mean free path
$L_\mathrm{ph}^\mathrm{ie}$ from our numerical calculations and
obtained that it can be described by a slowly variable function of
$\theta=\Tp /T$ which we denote by $F(\theta)$.
Its density dependence has been approximated by replacing the sum
over reciprocal lattice vectors $\bm g$ in the asymptotic expression
(\ref{Lambda-low-T}) with the integral over the vectors $\bm{g}$
uniformly distributed within the spherical layer of the inner radius
$\qBZ$ and the outer radius $2\,\kF $.
Then,
the Coulomb logarithm $\widetilde\Lambda^\mathrm{U}_\mathrm{e}$ has
been adjusted to the mean free path $L_\mathrm{ph}^\mathrm{ie}$
using Eqs.\ (\ref{kappa_phe0}), (\ref{S_col_e}), and
(\ref{kappa_ie_approx}).
 While integrating, we have used the
Debye-Waller factor which corresponds to a given temperature. In
analogy with the approximation of the electron thermal conductivity
by \citet{Gnedin}, the effect of the nucleus form factor has been
taken into account by introducing into the integrand an additional
factor $\exp\left(w_\mathrm{form}\,g^2/4\,\kF ^2\right)$. As a
result, the mean free path has been approximated as
\begin{eqnarray} \label{Fit}
  L_\mathrm{ph}^\mathrm{ie}&=&
  \frac{3}{2}\,\frac{m_\mathrm{i}\,\op \,\pF ^2\, \vF}
                     {Z^2e^4\,m_\mathrm{e}^\ast\,\qBZ \,n_\mathrm{e}}
               \,\frac{F(\theta)}{\Lambda_\mathrm{ph\,e}}
               \cr\cr
  &\approx& \frac{320\, \ai }{\left(1+\xr ^2\right) \Lambda_\mathrm{ph\,e}}
  \,\frac{26}{Z}\,
  \frac{F(\theta)}{0.01}
  \left(\frac{A\,\rho_6}{A^\prime}\right)^{1/2},
\end{eqnarray}
where
\begin{eqnarray}
    2\,\Lambda_\mathrm{ph\,e}&=&E_1(w\,y^2)-E_1(w)-\frac{\vF^2}{c^2}
    \frac{\mathrm{e}^{-w\,y^2}-\mathrm{e}^{-w}}{w}~,
   \\
    F(\theta)&=&0.014+\frac{0.03}{\exp(\theta/5)+1},
\end{eqnarray}
%
$w=w_\mathrm{DW}+w_\mathrm{form}$, and $y=\qBZ /(2\kF
)=(4Z)^{-1/3}$. The quantity
\begin{equation}
    w_\mathrm{DW}=1.683\,\sqrt{\frac{\xr }{
    A Z}}\,
    \left[0.5\,u_{-1}\,\exp(-9.1/\theta)+u_{-2}/\theta\right]
\end{equation}
measures the efficiency of the Debye-Waller factor
(\citealt{Baiko1995}); $w_\mathrm{form}$ determines the importance
of the form factor of atomic nuclei (see, e.g., \citealt{Gnedin});
$u_{-1}\approx 2.8$ and $u_{-2}\approx 13.0$ are dimensionless
phonon frequency moments (e.g., \citealt{PH73});
$E_1(x)=\int_x^\infty y^{-1}\exp(-y)\,{\rm d} y$ is the standard
exponential integral (whose approximations are given by
\citealt{AS}).

Now let us employ the ground-state composition of crustal matter and
the corresponding smooth composition model to describe proton charge
distribution within atomic nuclei in the crust (see Sect.\
\ref{sect:plasma} and Appendix B of \citealt{BOOK}). The radial
dependence of the proton number density within the proton core of
the nucleus ($r<R_\mathrm{p}$)
 is
 $n_\mathrm{p}(r) \propto \left[1- (r/R_\mathrm{p})^{t_\mathrm{p}}\right]^3$,
where $R_\mathrm{p}$
 and $t_\mathrm{p}$ are density dependent parameters.
In order to describe the effect of the nuclear form factor
it is sufficient to set
\begin{equation}
    w_\mathrm{form}=43\,x_\mathrm{nuc}^2,
\end{equation}
where
\begin{equation}
   x^2_\mathrm{nuc}=\frac{R_\mathrm{p}^2}{\ai^2}\frac{1-15/(5+t_\mathrm{p})
   +15/(5+2t_\mathrm{p})-5/(5+3t_\mathrm{p})}
   {1-9/(3+t_\mathrm{p})+9/(3+2t_\mathrm{p})-1/(1+t_\mathrm{p})}
\end{equation}
(\citealt{BOOK}, Appendix B).
Notice that the nucleus form factor and the associated parameter
$w_\mathrm{ form}$ are important only in inner crust, where the
phonon conductivity is rather insignificant. In the outer crust, one
can set $w_\mathrm{ form}=0$.

We have calculated $\kappa_\mathrm{ie}$ on a dense grid of $T$ and
$\rho$ values ($\log_{10} T~[\mathrm{K}]$ from 5 to 9, with the step
of 0.2; $\log_{10} \rho~[\mathrm{g~cm}^{-3}]$ from 5 to 14, with the
step of 0.1) restricted by the condition $\Gamma\ge 30$. The root
mean square relative fit error is 8\%, and the maximum error
$\approx 25$\% takes place at $T=6\times 10^7$ K
and
$\rho= 4\times
10^{12}$~g\,cm$^{-3}$. Had we restricted ourselves
to
the outer crust ($\rho\leq 4.3\times10^{11}$~g\,cm$^{-3}$), where
the phonon thermal conductivity can be really significant, then the
root mean square relative error would be 7\%, and the maximum error
of 21\% would occur at $T=4\times 10^7$ K and $\rho=2.5\times
10^{11}$~g\,cm$^{-3}$.
 For $\Gamma<\Gamma_\mathrm{ m}\approx 175$
 (the shaded region in Fig.\ \ref{fig:t-cond})
ions form a strongly non-ideal Coulomb liquid, and the
crystal-phonon description is, strictly speaking, invalid.
Nevertheless, as far as transport properties are concerned, a
strongly coupled  Coulomb liquid has much in common with a Coulomb
crystal (see, e.g., \citealt{Schmidt1997}, \citealt{Baiko1998},
\citealt{BOOK}). Therefore, we have extended calculations (and
included them in the fit) to $\Gamma=30$, because we expect that for
$30\lesssim \Gamma<175$ the adopted formalism still gives correct
order-of-magnitude estimates of the phonon-electron thermal
conductivity in the ion liquid (which is the thermal conductivity of
strongly coupled ions). Let us remind, that the electron transport
coefficients behave smoothly near the melting point (e.g.,
\citealt{Potekhin_e_cond}).

In fact, near the melting point (at $\Gamma\lesssim 175$)
phonon-electron scattering becomes unimportant for
ion conduction (see Fig.\ \ref{fig:diag} and Table \ref{tab:regimes})
and an uncertainty of our approximation of $\kappa_\mathrm{ie}$
does not affect the accuracy of the evaluation of the total
ion conductivity $\kappa_\mathrm{i}$.
A formal extension of our fit expression (\ref{Fit}) to
low $\Gamma\lesssim 30$ leads to
an unrealistic
exponential freezing of
the
ion-electron scattering (an exponential growth of
$\kappa_\mathrm{ie}$), which does not affect $\kappa_\mathrm{i}$
determined by ion-ion scattering.

Let us remark that if we neglect the Debye-Waller factor
($w_\mathrm{DW}=0$) and the nuclear form factor
($w_\mathrm{form}=0$), then we obtain
\begin{equation}
 \Lambda_\mathrm{ph\,e}\approx \,
 \ln\left(2\,\kF \over \qBZ \right)
         -\frac{\vF^2}{2c^2}\,
         \left(1-\frac{\qBZ ^2}{4\,\kF ^2}\right).
\end{equation}
Accordingly, $\Lambda_\mathrm{ph\,e}$ can be treated as a familiar
Coulomb logarithm (which takes into account a suppression of
backscattering for relativistic electrons) with the minimum electron
momentum transfer of $\qBZ $ and the maximum momentum transfer of
$2\,\kF $.

\section{Anisotropy of ion heat conduction}
\label{sect:IonMag}

Let us estimate magnetic field strengths which make
ion heat transport anisotropic. In a weakly coupled Coulomb
plasma ($\Gamma\ll 1 $) the effect of the magnetic field
on the ion thermal conductivity is well known  (e.g.,
\citealt{Braginski1963}). Specifically, the magnetic
field does not affect noticeably the ion transport
as long as the ion magnetization parameter is small,
$\omega_{B\mathrm{i}}\tau_\mathrm{i} \lesssim 1$, where $\omega_{B \rm
i}=Z e B/(m_\mathrm{i}c)$ is the ion cyclotron frequency.

In a Coulomb crystal or liquid, we expect that the magnetic field
strongly influences ion heat transport provided it distorts the
spectrum of heat carrying phonons. In a classical crystal ($T\gtrsim
\Tp /3$) the heat is carried by all phonons whose spectrum is
distorted when  $\omega_{B\mathrm{i}} \gtrsim \op $
(\citealt{Baiko2000,BOOK}, \S~4.1.6b). The latter inequality will be
treated as a
condition for an  anisotropic
ion thermal conductivity; we will use it for ion liquid and solid.
In the quantum crystal ($T\ll T_\mathrm {p}$) heat is mainly
transported by low-frequency phonons (Sec.\ \ref{sect:U-low-T}),
with $\omega \sim \omega_T\sim \kB T/\hbar$. Their spectrum is
affected by the magnetic field at $\omega_{B\mathrm{i}}\gtrsim
\omega_T$ (\citealt{Baiko2000,BOOK}, \S 4.1.6b). Accordingly, the
criterion for anisotropic ion conduction can be written as
$\omega_{B\mathrm{i}}
\gtrsim
\op \,T/\Tp $. The same criteria can be applied for describing the
effects of the magnetic field on
the
thermodynamic properties of Coulomb crystal. They
are in a qualitative agreement with the results of accurate
calculations
 (\citealt{Baiko2000,BOOK}, \S 4.1.6b, their Fig.\
4.3), which can serve as an additional argument in favor of our criterion.

Basing on the above estimates we can introduce a characteristic
magnetic field $B_\mathrm{m}$, which starts to affect ion
heat transport and makes this transport anisotropic.
At $B \lesssim B_\mathrm{m}$ one can neglect the effect
of the magnetic field on $\kappa_\mathrm{i}$ and
use the results of preceding sections. We can suggest
a simple estimate of $B_\mathrm{m}$; it is valid in a wide
range of densities and temperatures including all
values of $\rho$ and $T$ shown in Fig.\ \ref{fig:diag},
%
\begin{eqnarray}
B_\mathrm{m}&=&\frac{m_\mathrm{i}c}{Ze}\,
      \left(
        \tau_\mathrm{ii}^2
        +\frac{1}{\op ^2}
        +\frac{1}{\omega_T^2}
      \right)^{-1/2}
    \approx
    10^{14}\,\left(\rho_6\,A/A'\right)^{1/2}\,
     \nonumber \\
     &\times&
    \left[
        3\pi\,\Gamma^{-3}/\Lambda_\mathrm{ii}^2+1+\left(\Tp /T\right)^2
    \right]^{-1/2}\,\mathrm{G},
    \label{B_m}
\end{eqnarray}
%
%
where $\tau_\mathrm{ii}$ is given by Eq.\ (\ref{tau-gas}) with
the Coulomb logarithm
$\Lambda_\mathrm{ii}=\ln\left[2+1/(\sqrt{3}\Gamma^{3/2})\right]$.
The magnetic field $B \gtrsim10^{14}\,\rho_6^{1/2}$~G,  for which
$\omega_{B\mathrm{i}}\gtrsim\op $, strongly affects ion heat
transport at any temperature. At lower field $B$, the condition $B
\sim B_\mathrm{m}$ is realized at sufficiently low or sufficiently
high temperatures. In the case of low temperatures, a magnetized
Coulomb crystal possesses very soft phonon modes which are
responsible for heat transport and most sensitive to the magnetic
field (\citealt{Baiko2000,BOOK}, \S 4.1.6b, Fig.\ 4.3).  In the high
temperature plasma, the ions become weakly coupled and their
relaxation time (\ref{tau-gas}) rapidly increases with the growth of
$T$, increasing a typical rotation angle of ions between successive
collisions, and hence the magnetic field effect.

The boundaries of domains, where $B \gtrsim B_\mathrm{m}$
and $\kappa_\mathrm{i}$ is anisotropic, are shown in Fig.\ \ref{fig:diag}
for $B_\mathrm{m}=10^{13}$ and
$B_\mathrm{m}=10^{14}$~G. The domains themselves
are situated below and to the left of these boundaries.
A not very strong field $B=3\times10^{12}$~G
(employed in Figs.\
\ref{fig:t-cond} and \ref{fig:anisotrop})
does not affect ion conduction at all $T$ and $\rho$
in Fig.\ \ref{fig:diag}.

Let us notice that at rather low densities $\rho\lesssim
\rho_{B\mathrm{e}}$ (see Sec.\ \ref{sect:kappae}) the
plasma electrons occupy
only one or several Landau levels. This circumstance can strongly
modify electron-phonon scattering. We have neglected this effect in
the present publication. However, at these low densities the ion
thermal conductivity (at not too low temperatures $T\gtrsim 10^6 K$)
is mainly determined by ion-ion scattering (see Fig.\
\ref{fig:diag}) and electron-phonon scattering is relatively
unimportant. At high densities the electrons populate many Landau
levels so that the Landau level structure can be neglected as
confirmed by calculations of electron transport properties
\citep{Potekhin_magn}.

\section{Discussion}
\label{sect:discuss}

\begin{figure}
    \centering \resizebox{3.12in}{!}{\includegraphics[]{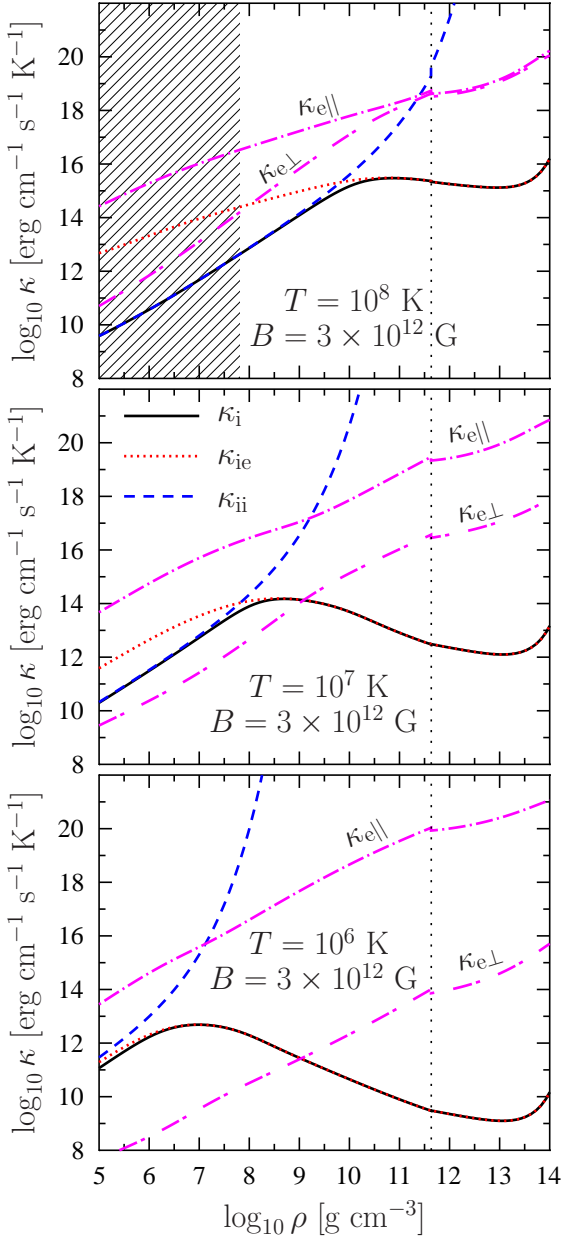}}
    \caption{(Color online) Thermal conductivity of the ground-state
        neutron star crust
        for
        $T=10^6$, $10^7$, and $10^8$~K. The solid line is the total
        ion (phonon)
        conductivity $\kappa_\mathrm{i}$.
        The dash and dot lines show
        $\kappa_\mathrm{ie}$ and
        $\kappa_\mathrm{ii}$,  respectively. The dot-dash lines are the  electron
        thermal conductivities,  across and along the magnetic field
        $B=3\times 10^{12}$~G, calculated using a simplified model
        of Sect.\ \ref{sect:kappae}. The vertical dotted line shows the neutron
        drip density $\rho_\mathrm{ND}=4.3\times 10^{11}$~g~cm$^{-3}$.
        The shaded zone in the upper panel corresponds  to
        a strongly coupled Coulomb liquid ($T>T_\mathrm{m}$, see
        Fig.\ \ref{fig:diag}). At $T=10^7$ and
        $10^6$~K the ions form a crystal for all displayed densities.
    }
    \label{fig:t-cond}
\end{figure}

Let us compare the ion thermal conductivity (calculated neglecting
the effects of the magnetic field, as discussed in Sect.\
\ref{sect-B}) with the electron thermal conductivities
(\ref{kappae}) along and across the magnetic field $\bm{B}$.

Figure \ref{fig:t-cond} shows the density dependence of the partial
ion conductivities $\kappa_\mathrm{ii}$ and $\kappa_\mathrm{ie}$,
the total ion conductivity $\kappa_\mathrm{i}$, and also of the
electron conductivities along and across $\bm{B}=3\times
10^{12}~{\rm G}$, $\kappa_\mathrm{e\parallel}$ and
$\kappa_\mathrm{e\perp}$, at three values of $T=10^6$, $10^7$ and
$10^8$~K. One sees that the longitudinal electron thermal
conductivity $\kappa_\mathrm{e
\parallel}$ always dominates over the thermal conductivity of ions.
Therefore, the ion conductivity cannot contribute into heat
transport along magnetic field lines under typical conditions in
neutron star envelopes. Nevertheless, it can compete with
$\kappa_\mathrm{e \perp}$ and be the dominant thermal conductivity
across the magnetic field lines, especially at not too high
densities (in the outer neutron star envelope) and temperatures. For
instance, at $B=3\times 10^{12}$~G and $T=10^8$~K (the upper panel
in Fig.\ \ref{fig:t-cond}) $\kappa_\mathrm{i}$ does not dominate in
the transverse conduction at all, but at $T=10^{6-7}$~K (the middle
and bottom panels) it dominates in the outer layer at densities
$\rho \lesssim 10^9$~g~cm$^{-3}$. This density range is most
important in the neutron star physics because it belongs to the
neutron-star heat blanketing envelope which shields warm neutron
star interior from the efficient cooling via thermal conduction to
the surface and then through the thermal emission from the surface
(see, e.g., \citealt{BOOK}). Higher magnetic fields would stronger
suppress $\kappa_\mathrm{e \perp}$ and widen the density range where
ion thermal conduction across $\bm{B}$ dominates over electron one.

It is important to emphasize the efficiency of phonon-electron
scattering (neglected by \citealt{Perez2006}). This scattering is
most efficient and determines $\kappa_\mathrm{i}$ at sufficiently
high densities and low temperatures (in domains IV and V in Fig.\
\ref{fig:diag}), while phonon-phonon (ion-ion) scattering overtakes
ion transport at lower densities and higher temperatures (in domains
I--III). Phonon-electron scattering reduces the total ion
conductivity $\kappa_\mathrm{i}$ in comparison with the conductivity
$\kappa_\mathrm{ii}$ that is solely determined by phonon-phonon
scattering. In particular, for $B=3 \times 10^{12}$~G and $T=10^8$~K
(the upper panel in Fig.\ \ref{fig:t-cond}) this suppression is very
efficient at $\rho \gtrsim 3 \times 10^{10}$ g~cm$^{-3}$.
When the star cools, phonon-electron
scattering becomes more important, and the layer of its dominance
expands to lower $\rho$. For instance, at $T=10^6$~K (the bottom
panel of Fig.\ \ref{fig:t-cond}) it dominates at all densities (see
also Fig.\ \ref{fig:diag}).

\begin{figure}
 \centering \resizebox{3.2in}{!}{\includegraphics[]{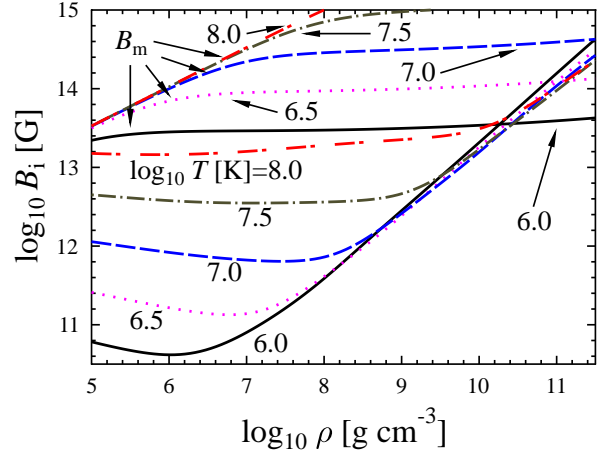}}
 \caption{(color online) Characteristic magnetic field $B_\mathrm{i}$,
    which makes the ion thermal conductivity
    $\kappa_\mathrm{i}$ larger than the
    transverse electron thermal conductivity $\kappa_{e \perp}$,
    and characteristic field $B_\mathrm{m}$, which makes
    ion thermal conduction anisotropic,
    versus density, for five values
    of the temperature with $\log_{10}T[K]=6.0$,
    6.5, 7.0, 7.5 and 8.0.
 }
 \label{fig:B_cr}
\end{figure}

If we fix $\rho$ and $T$ and increase $B$ then the heat transport
across $\bm{B}$ will be determined by ions
($\kappa_\mathrm{i}\gtrsim \kappa_\mathrm{e \perp}$) after $B$
exceeds some value $B_\mathrm{i}$. Figure \ref{fig:B_cr} shows the
density dependence of $B_\mathrm{i}$. At low $\rho$, this density
dependence is weak but $B_\mathrm{i}$ is a strong function of the
temperature (noticeably decreases with decreasing $T$). At large
$\rho$, the field $B_\mathrm{i}$ becomes almost independent of $T$
(all curves in Fig.\ \ref{fig:B_cr} merge into nearly one curve) but
increases with growing $\rho$. For instance, one needs an ordinary
pulsar magnetic field $B \gtrsim 10^{12}$~G for the ion conduction
to dominate over $\kappa_\mathrm{e \perp}$ at $\rho=10^9$
g~cm$^{-3}$ (for $T \lesssim 10^7$~K). However, the neutron star
must have a typical magnetar field $B \gtrsim 10^{14}$~G to reach
similar dominance at $\rho \sim 10^{11}$ g~cm$^{-3}$ (for all
temperatures in Fig.\ \ref{fig:B_cr}). Also, in Fig.\ \ref{fig:B_cr}
we show the values of the magnetic fields, $B_\mathrm{m}$, which
make ion heat transport anisotropic for the same temperatures $T$.
At not too high densities, $\rho\lesssim 10^{10}$~g~cm$^{-3}$, the
ions become the leading heat carriers across the magnetic fields $B
\lesssim B_\mathrm{m}$. However, at $\rho \sim 10^{11}$~g~cm$^{-3}$
the magnetic field $B \gtrsim 10^{14}$~G, which makes the transverse
ion heat transport dominant, becomes larger than $B_\mathrm{m}$ if
the temperatures falls below $T\sim 10^6$~K. However, even in that
case it would be preferable to use the ion thermal conductivity at
$B=0$ than to neglect the ion thermal conductivity at all.

\begin{figure}
\centering
\resizebox{3.2in}{!}{\includegraphics[]{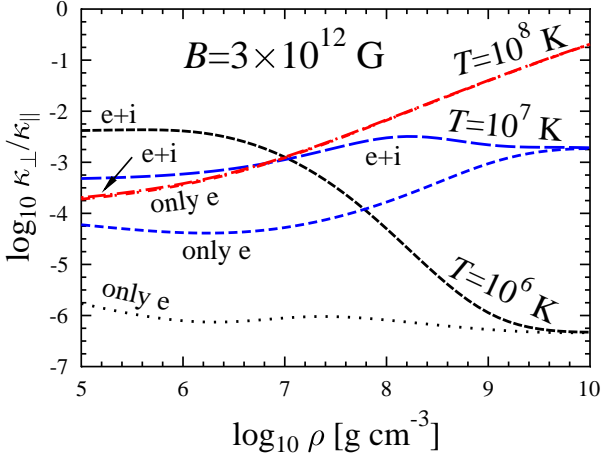}}
\caption{(color online) Anisotropy of the thermal
conductivity
  $\kappa_\perp/\kappa_\parallel$ versus density for
  the magnetic field $B= 3\times10^{12}$~G
  and three values of the temperature
  $\log_{10}T[K]$=6.0, 7.0, and 8.0. See text for details.}
\label{fig:anisotrop}
\end{figure}

A magnetic field introduces a considerable anisotropy in thermal
conduction of the neutron star envelope. For illustration, in Fig.\
\ref{fig:anisotrop} we plot the density dependence of the
anisotropy, which is the ratio of the transverse conductivity
$\kappa_\perp=\kappa_\mathrm{i}+\kappa_\mathrm{e \perp}$ to the
longitudinal conductivity
$\kappa_\parallel=\kappa_\mathrm{i}+\kappa_\mathrm{e
\parallel}\approx \kappa_\mathrm{e \parallel}$ of the matter.
We take the same magnetic field $B= 3 \times 10^{12}$~G and the same
three values of the temperature $\log_{10}T$=6, 7, and 8~K, as in
Fig.\ \ref{fig:t-cond}. The lines marked `e+i' are plotted
taking
into account both, electron and ion, conductivities. We see that the
anisotropy can be very strong, and it increases with lowering $T$,
when the star cools. For instance, we have
$\kappa_\perp/\kappa_\parallel \sim 0.1$ for $T=10^8$~K and $\rho
\sim 10^8$~g~cm$^{-3}$ but $\kappa_\perp/\kappa_\parallel \sim
10^{-6}$ for $T=10^6$~K and the same $\rho \sim 10^8$ g~cm$^{-3}$.
To emphasize the importance of the ion conduction, the lines marked
as `only e' in Fig.\ \ref{fig:anisotrop} show the anisotropy
$\kappa_\mathrm{e \perp}/\kappa_\mathrm{e \parallel}$ produced by
electron conduction alone. At $T=10^8$~K the ion conductivity is
small and does not affect heat transport as discussed above (curves
`e+i' and `only e' merge). However, at lower $T$ ion conduction
becomes important. It is clearly seen that $\kappa_\mathrm{i}$
greatly reduces the heat conduction anisotropy at sufficiently low
$\rho$, where ion conduction dominates in the transverse heat
transport (see Fig.\ \ref{fig:t-cond} and the discussion above). For
instance, taking $T=10^6$~K and $\rho=10^6$ g~cm$^{-3}$ and
neglecting the ion contribution, we would have a very strong heat
transport anisotropy $\kappa_\mathrm{e \perp}/\kappa_\mathrm{e
\parallel}\approx 10^{-6}$, while the inclusion of the ion
conductivity gives
$\kappa_\mathrm{\perp}/\kappa_\mathrm{\parallel}\sim 10^{-3}$. This
means that the heat transport becomes much more isotropic (by
approximately three orders of magnitude). This effect may strongly
affect the temperature distribution in the neutron star heat
blanketing envelope and over the surface, the relation between the
internal neutron star temperature and surface thermal luminosity,
and the neutron star cooling.

\section{Conclusions}
\label{sect:conclusions}

We have calculated the thermal conductivity of ions
$\kappa_\mathrm{i}$ in large ranges of the density and temperature
in a neutron star envelope. We have taken into account ion-ion and
ion-electron scattering (equivalent to phonon-phonon and
phonon-electron scattering at sufficiently low temperatures) and
analyzed various ion-conduction regimes. We have performed extensive
Monte Carlo calculations of $\kappa_\mathrm{ie}$ due to
phonon-electron scattering. All our results are approximated by
simple analytic expressions. In this way we have got a systematic
description of all ion conduction regimes in neutron star envelopes.
We have compared calculated values of $\kappa_\mathrm{i}$ with the
values of the electron thermal conductivity $\kappa_\mathrm{e}$
which is usually the only thermal conductivity taken into account in
neutron star envelopes (except for neutron star atmospheres, where
heat transport is radiative).

Our main conclusions are:

\begin{enumerate}

\item
The ion thermal conductivity is
much less affected
by neutron star magnetic fields than the electron
thermal conductivity (and we have neglected the effects
of magnetic fields on $\kappa_\mathrm{i}$).

\item
The ion thermal conductivity is typically much lower
than the electron thermal conductivity along
magnetic field lines in the neutron star envelope.
Heat conduction along the magnetic field lines is
mostly provided by electrons.

\item
The ion thermal conductivity can be higher than the electron thermal
conductivity across magnetic field lines in the outer envelope of a
neutron star at not too high temperatures and densities ($T \lesssim
10^8$~K and $\rho \lesssim 10^9$ g~cm$^{-3}$ for $B \sim 3 \times
10^{12}$~G).

\item
The conductivity $\kappa_\mathrm{i}$ at low temperatures and high
densities is determined by electron-ion collisions (e.g., at $\rho
\gtrsim 3 \times 10^8$ g~cm$^{-3}$ for $T \sim 10^7$~K) although at
lower $\rho$ and higher $T$ it is determined by ion-ion collisions.

\item
The inclusion of
$\kappa_\mathrm{i}$ can strongly
(by a few orders of magnitude) reduce
large anisotropy of thermal conduction in magnetized
neutron star envelopes, affecting thus temperature
distribution in the heat blanketing envelope and
over the stellar surface, and cooling of magnetized
neutron stars.

\end{enumerate}

Our results extend those obtained by \cite{Perez2006}. First, we
have included the contribution of electron-phonon scattering.
Second, we have improved the consideration of phonon-phonon
scattering. Finally, we have analyzed different ion conduction
regimes. Nevertheless, we stress that our results are
semi-quantitative and can be elaborated further. In particular, it
would be interesting to consider phonon scattering in impure and
imperfect crystals (by studying phonon-impurity scattering neglected
here), to take into account electron band structure effects in
phonon-electron scattering, to describe  accurately phonon-phonon
scattering [without using the estimate Eq.\ (\ref{cl-latt})],
 and the
effects of magnetic fields on the ion thermal conductivity.
All these problems open
a new and interesting field
of neutron star kinetics which goes far beyond the scope
of the present paper. We expect to deal with them
in future publications.

\section*{Acknowledgments}
  We are deeply grateful to D.G.\ Yakovlev,
who suggested the topic of the present paper, and whose helpful
comments and guidance through the difficult problems of kinetics of
dense matter were crucial for completing this study. We are grateful
A.~Y. Potekhin and  Yu.~A.\ Shibanov for useful comments and
discussions and to the referee, U.R.M.E. Geppert, for critical
remarks. This work was partially supported by the Polish MNiI grant
No.\ 1P03D.008.27, the Russian Foundation for Basic Research (grants
05-02-16245, 05-02-22003) by the Federal Agency for Science and
Innovations (grant NSh 9879.2006.2) and a grant of the Dynasty
Foundation and the International Center for Fundamental Physics in
Moscow.


\label{lastpage}

\begin{thebibliography}{}
%
\bibitem[Abramowitz \& Stegun(1972)]{AS}
Abramowitz M., Stegun I.A. (eds.), 1972, Handbook of Mathematical
Functions, Dover, New York

\bibitem[Baiko \& Yakovlev(1995)]{Baiko1995}
Baiko D.A., Yakovlev D.G., 1995, Astron.\
 Lett.\ 21, 702

\bibitem[Baiko et al.(1998)]{Baiko1998}
Baiko D.A., Kaminker A.D., Potekhin A.Y., Yakovlev D.G., 1998,
Phys.\ Rev.\ Lett., 81, 5556


\bibitem[Baiko(2000)]{Baiko2000}
Baiko D.A., 2000, ''Kinetic phenomena in cooling neutron stars'',
PhD thesis (Ioffe Phys.-Tech. Inst., St. Petersburg) [in Russian],
unpublished

\bibitem[Baiko et al.(2001)]{Baiko2001}
Baiko D.A.,  Potekhin A.Y., Yakovlev D.G., 2001, Phys.\ Rev.\ E, 64,
057402

\bibitem[Baiko(2002)]{Baiko2002}
Baiko D.A., 2002, Phys.\ Rev.\ E, 66, 056405


\bibitem[Bernu \& Vieillefosse(1978)]{BV78}
Bernu B., Vieillefosse P., 1978, Phys.\ Rev.\ A, 18, 2345

%
\bibitem[Burwitz et al.(2003)]{Burwitz2003}
    Burwitz V., Haberl F., Neuh\"{a}user R., Predehl P., Tr\"{u}mper J.,
     Zavlin V.E., 2003, A \& A, 399, 1109


\bibitem[\protect\citeauthoryear{Braginski}{1963}]{Braginski1963}
  Braginski S.I., 1963, in
  Voprosy Teorii Plasmy, vol.1, ed. by M.A. Leontovich
  (GosAtomIzdat: Moscow); English translation: 1965,
  Reviews of Plasma Physics, vol.1, ed. by M.A. Leontovich,
  (Consultants Bureau: New York) p. 205

\bibitem[\protect\citeauthoryear{Cohen \& Keffer}{1955}]{CohenKeffer55}
Cohen M.H., Keffer F., 1955, Phys.\ Rev., 99, 1128

\bibitem[\protect\citeauthoryear{Daligault}{2006}]{Hopping}
Daligault J., Phys.\ Rev.\ Lett., 96, 065003

\bibitem[\protect\citeauthoryear{Dubin}{1990}]{Dubin90}
Dubin D.H.E., Phys.\ Rev.\ A, 42, 4972

\bibitem[\protect\citeauthoryear{Gepert, Kueker \& Page}
{Geppert et al.}{2004}]{Geppert2004}
    Geppert U., Kueker M., Page D., 2004,
    A \& A, 426, 267

\bibitem[\protect\citeauthoryear{Gepert, Kueker \& Page}
{Geppert et al.}{2006}]{Geppert2006}
    Geppert U., Kueker M., Page D., 2006,
    A \& A, 457, 937

\bibitem[\protect\citeauthoryear{Gnedin, Yakovlev \& Potekhin}
{Gnedin et al.}{2001}]{Gnedin}
    Gnedin O.Y., Yakovlev D.G.,
    Potekhin A.Y., 2001, MNRAS, 324, 725

\bibitem[\protect\citeauthoryear{Flowers \& Itoh}{1976}]{FI76}
Flowers~E., Itoh~N., 1976,
   ApJ, 206, 218

\bibitem[\protect\citeauthoryear{Haberl}{2007}]{Haberl2007}
Haberl F., Astrophys Space Sci., 308, 181

\bibitem[\protect\citeauthoryear{Haensel, Potekhin \& Yakovlev}
{Haensel et al.}{2007}]{BOOK} Haensel P., Potekhin A.Y., and
Yakovlev D.G., 2007, Neutron Stars 1: Equation of State and
Structure. Springer Verlag, New York

\bibitem[\protect\citeauthoryear{Ho}{2007}]{Ho2007}
Ho W.C.G., 2007, MNRAS, 380, 71


\bibitem[\protect\citeauthoryear{Landau \& Lifshitz}{1993}]{Stati1}
Landau L.D., Lifshitz E.M., 1993, Statistical Physics, Part 1.
Pergamon, Oxford

\bibitem[\protect\citeauthoryear{Lifshitz \& Pitaevski}{1980}]{Stati2}
Lifshitz E.M., Pitaevski\u{\i}  L.P., 1980, Statistical Physics,
Part 2. Pergamon, Oxford


\bibitem[\protect\citeauthoryear{McGaughey \& Kaviany}{2006}]{Ions_to_phons}
McGaughey A.J.H., Kaviany M., 2006, Advances in Heat Transfer, 39,
169


\bibitem[\protect\citeauthoryear{Negele \& Vautherin}{1973}]{NV}
 Negele~J.W., Vautherin~D., 1973,
      Nucl.~Phys.~A, 207, 298

\bibitem[\protect\citeauthoryear{Oyamatsu}{1993}]{Oyamatsu} Oyamatsu~K.,
    Nucl.~Phys.~A, 561, 431

\bibitem[\protect\citeauthoryear{Raikh \& Yakovlev}{1982}]{RaikhYakovlev}
  Raikh M.E., Yakovlev D.G., 1982, Astrophys.\ Sp.\
   Sci., 87, 193

\bibitem[\protect\citeauthoryear{Page, Geppert \& Weber}
{Page et al.}{2006}]{Page2006}
 Page D., Geppert U., Weber F., 2006,
  Nucl.\ Phys.\  A 777, 497

\bibitem[\protect\citeauthoryear{P\'erez-Azor\'in, Miralles J.A. \& Pons J.A.}
{P\'erez-Azor\'in et al.}{2006}]{Perez2006} P\'erez-Azor\'in J.F.,
Miralles J.A., Pons J.A., 2006, A\&A, 451, 1009.

\bibitem[\protect\citeauthoryear{Pethick \& Thorsson}{1997}]{PT97}
Pethick C.J., Thorsson V., 1997, Phys.\ Rev.\ D, 56, 7548


\bibitem[\protect\citeauthoryear{Pierleoni, Ciccotti \& Bernu}
{Pierleoni et al.}{1987}] {Pierleoni1987}
  Pierleoni C., Ciccotti G., Bernu B., 1987, Europhysics Letters,
  {\bf 4}, 1115

\bibitem[\protect\citeauthoryear{Potekhin, Chabrier \&
Yakovlev} {Potekhin et al.}{1997}]{Potekhin97} Potekhin A.Y.,
Chabrier G., Yakovlev D.G., 1997, A\&A 323, 415

\bibitem[\protect\citeauthoryear{Potekhin}{1999}] {Potekhin_magn}
 Potekhin A.Y., 1999, A\&A, {\bf 351}, 787

\bibitem[\protect\citeauthoryear{Potekhin et al.}{1999}]{Potekhin_e_cond}
Potekhin A.Y., Baiko D.A., Haensel P., Yakovlev D.G., 1999, A\&A,
346, 345.

\bibitem[\protect\citeauthoryear{Pollock \& Hansen}{1973}]{PH73}
Pollock L.E., Hansen J.P., 1973, Phys.\ Rev.\ A, 8, 3110



\bibitem[\protect\citeauthoryear{Schmidt et al.}{1997}]{Schmidt1997}
Schmidt P., Zwicknagel G., Reinhardt P.G., Toepffer C., 1997,
 Phys.\ Rev.\ E, 56, 7310


\bibitem[\protect\citeauthoryear{Shternin \& Yakovlev}{2006}]{Shternin} Shternin P.S.,
Yakovlev D.G., 2006, Phys.\ Rev.\ D, 74, 043004

\bibitem[\protect\citeauthoryear{Ventura \& Potekhin}{2001}]{VP}
Ventura J., Potekhin A.Y., 2001,
in The Neutron Star -- Black Hole Connection, NATO Science Ser. C,
567, edited by C. Kouveliotou, E.P.J. van den Heuvel, \& J. Ventura
(Kluwer, Dordrecht), 393


\bibitem[\protect\citeauthoryear{Yakovlev \& Kaminker}{1994}] {YK}
Yakovlev D.G., Kaminker A.D., 1994
in The Equation of State in Astrophysics, edited by G. Chabrier \&
E. Schatzman (Cambridge University Press, Cambridge), 214

\bibitem[\protect\citeauthoryear{Ziman}{1960}]{Ziman}
Ziman J.M., 1960, Electrons and Phonons. Oxford Univ.\ press, Oxford





\end{thebibliography}
\end{document}